\documentclass[twocolumn,trackchangess]{aastex701}
\usepackage{graphicx}
\usepackage{epstopdf}
\usepackage[utf8]{inputenc}
\usepackage{hyperref}  

\usepackage{soul}
\soulregister\citeyear7
\soulregister\citep7
\soulregister\citet7
\usepackage{xcolor}
\newcommand{\revone}[1]{#1}

\DeclareRobustCommand{\lhlsout}{\bgroup\markoverwith{\textcolor{red}{\rule[.5ex]{2pt}{0.4pt}}}\ULon}

\usepackage{booktabs}
\usepackage{threeparttable}
\usepackage{makecell}
\usepackage{adjustbox}
\usepackage{amsmath}

\begin{document}

\title{Mass–Radius Constraints for 2S 0918–549 from an RXTE Superexpansion Burst: A Direct Cooling-Tail Analysis}
\correspondingauthor{Helei Liu}
\email{heleiliu@xju.edu.cn}

\author[sname='Fan']{Hongbin Fan}
\affiliation{School of Physical Science and Technology, Xinjiang University, Urumqi 830046, China}
\email{}  

\author[orcid=0000-0001-8706-1882,gname=Helei, sname='Liu']{Helei Liu} 
\affiliation{School of Physical Science and Technology, Xinjiang University, Urumqi 830046, China}
\email{heleiliu@xju.edu.cn}

\author[orcid=0000-0003-2310-8105,gname=Zhaosheng,sname=Li]{Zhaosheng Li}
\affiliation{School of Science, Qingdao University of Technology, Qingdao 266525, China}
\email{lizhaosheng@xtu.edu.cn}

\author[orcid=0000-0001-8768-3294]{Yupeng Chen}
\affiliation{Key Laboratory for Particle Astrophysics, Institute of High Energy Physics, Chinese Academy of Sciences, 19B Yuquan Road, Beijing 100049, China}
\email{}

\author[sname=Ban,gname=Shoutao]{Shoutao Ban}
\affiliation{School of Physical Science and Technology, Xinjiang University, Urumqi 830046, China}
\email{}

\author[0000-0002-3839-4864]{Guoliang L\"{u}}
\affiliation{School of Physical Science and Technology, Xinjiang University, Urumqi 830046, China}
\email{}

\author[orcid=0000-0001-8726-5762]{Akira Dohi}
\affiliation{RIKEN Pioneering Research Institute (PRI), 2-1 Hirosawa, Wako, Saitama 351-0198, Japan}
\affiliation{RIKEN Center for Interdisciplinary Theoretical and Mathematical Sciences (iTHEMS), RIKEN 2-1 Hirosawa, Wako, Saitama 351-0198, Japan}
\email{}

\author{Chunhua Zhu}
\affiliation{School of Physical Science and Technology, Xinjiang University, Urumqi 830046, China}
\email{}

\author[0000-0002-9042-3044]{Renxin Xu}
\affiliation{Department of Astronomy, Peking University, Beijing 100871, China}
\affiliation{Kavli Institute for Astronomy and Astrophysics, Peking University, Beijing 100871, China}
\email{r.x.xu@pku.edu.cn}

\begin{abstract}
Thermonuclear (Type I ) X-ray bursts from accreting neutron stars offer a means to determine neutron-star (NS) mass ($M$) and radius ($R$) and thereby probe the properties of matter at supranuclear density. A subset of these events, photospheric radius–expansion (PRE) bursts, provide a particularly powerful tool to constrain the neutron-star $M$ and $R$. Here, we apply the direct cooling-tail method to 2S~0918$-$549, using a rare superexpansion burst observed by \emph{RXTE}. We fit only the post–touchdown data within \(F/F_{\rm td}\in[0.6,0.95]\), employing modern atmosphere models (pure He and metal-enriched). The pure-He atmosphere yields a good description of the cooling tail (\(\chi^{2}/\nu=18.12/14\)), whereas metal-rich models fail; information-criterion tests (AIC/BIC) disfavor adding a free absorption edge in every time bin, indicating that heavy-element ashes are unnecessary. The joint fit gives a distance \(d=4.1-5.3\) kpc and mass–radius constraints \(M=1-2\,M_\odot\) and \(R=9.7-11.9\) km (99\% confidence). These results suggest that representative families of both gravity-bound and self-bound equations of state remain viable at the $1\sigma$ confidence level.

\end{abstract}


\section{Introduction} \label{sec:intro}

Neutron stars are compact objects with densities second only to those of black holes, exceeding nuclear saturation density and producing conditions that cannot be reproduced in terrestrial laboratories. Measuring their macroscopic properties, most notably mass and radius, provides a direct probe of the neutron-star equation of state\citep{2012ARNPS..62..485L}. Precise neutron-star masses can be obtained from radio pulsar timing. In double neutron star systems, combining the mass function with post-Keplerian parameters enables high-precision mass determinations \citep{1992RSPTA.341..117T,2010ApJ...722.1030W,2016RvMP...88b1001W}. PSR~B1913+16 was the first neutron star whose mass was precisely measured by post-Keplerian timing parameters \citep{2010ApJ...722.1030W}; see also its discovery paper \citep{1975ApJ...195L..51H}. However, for many millisecond pulsars, the near-circularity of their orbits makes it difficult to measure multiple post-Keplerian effects, limiting mass constraints \citep{2016ARA&A..54..401O,2016RvMP...88b1001W}. Several complementary techniques have been developed to probe neutron star radii, including spectral fitting that combines observed fluxes with independent distances, pulse profile (waveform) modeling, and thermal-emission atmosphere modeling \citep{2006MNRAS.373..836P,2011A&A...527A.139S,2016EPJA...52...63M}. Nevertheless, precise and simultaneous constraints on both mass ($M$) and radius ($R$)—and thus on compactness (M/R)—remain challenging \citep{2016EPJA...52...63M}. Thermonuclear (Type I ) X-ray bursts provide an important alternative approach for joint mass-and-radius inference using their fluxes and spectra \citep{2008ApJS..179..360G}.

In low-mass X-ray binaries (LMXBs), neutron stars accrete H/He-rich material from their companions. As the accreted layer is compressed and heated on the stellar surface, unstable thermonuclear burning ignites once the critical temperature–density conditions are reached, producing Type I X-ray bursts with rise times of a few seconds and total radiated energies of order $10^{39}\,\mathrm{erg}$ over $\sim15$–$50$ s \citep{1993SSRv...62..223L}. In some bursts, radiation pressure balances gravity and lifts the photosphere above the stellar surface, halting the inflow from the inner disk; these photospheric radius-expansion (PRE) events exhibit characteristic signatures: the apparent blackbody radius expands to several times the neutron-star radius while the color temperature decreases. As the flux subsequently declines, the photosphere contracts back toward the stellar surface and the temperature increases again, reaching maximum at the so-called ``touchdown'' moment, when the photosphere has settled back onto the surface and the luminosity remains close to the Eddington limit \citep{1993SSRv...62..223L}.  

Most Type I bursts spectra are well described by blackbody emission, allowing time-resolved spectroscopy to track the evolution of the blackbody temperature $kT_{\rm bb}$ and the blackbody normalization $K_{\rm bb}$. PRE bursts display a distinctive spectral evolution in which, near touchdown, $kT_{\rm bb}$ reaches a local maximum while $K_{\rm bb}$ reaches a local minimum. Under the commonly adopted assumption that the bolometric flux at touchdown is close to the Eddington flux, this epoch enables joint mass–radius inference for neutron stars \citep{2009ApJ...693.1775O,2010ApJ...712..964G,2010ApJ...719.1807G,2012ApJ...748....5O}.
Several methods have been proposed and applied to determine neutron-star radii using thermonuclear (Type I) X-ray bursts. \cite{1979ApJ...234..609V} proposed that the apparent angular diameter measured during the cooling tails of bursts could be used to obtain joint mass–radius (M–R) constraints. Subsequently, different strategies to break the mass–radius degeneracy using multiple spectroscopic observables have been discussed \citep{1986ApJ...305..246F,1987MNRAS.226...39S,1990A&A...237..103D,2006Natur.441.1115O}.
\cite{1987PASJ...39..287E} further suggested estimating the gravitational redshift and stellar radius by tracking the luminosity–color-temperature evolution during PRE. 
\cite{1990A&A...237..103D} conducted a systematic “touchdown” flux analysis, refining $M/R$ estimates by identifying the moment when the photosphere returns to the stellar surface.

Comprehensive reviews of PRE-burst techniques were provided by \citet{1993SSRv...62..223L}.
Building on this framework, \cite{2012ApJ...748....5O} introduced a Bayesian inference to analyze touchdown fluxes and apparent areas in ensembles of time-resolved spectra.
\cite{2017MNRAS.466..906S} proposed that, after touchdown, the emitting area is consistent with the full stellar surface so that changes in the blackbody normalization mainly trace the evolving color-correction factor $f_{\rm c}\!=\!T_{\rm c}/T_{\rm eff}$; this underpins the cooling-tail fitting approach and its subsequent refinements.
In recent years, advances in NS-atmosphere models and multi-messenger constraints have significantly improved their precision, constraining the neutron star equation of state (EOS) \citep{2017A&A...608A..31N,2018PhRvL.121.161101A,2019ApJ...887L..21R,2019ApJ...887L..24M,2021ApJ...918L..27R,2021ApJ...918L..28M}.

With the growing sample of PRE bursts, a special subclass has emerged in which a millisecond–to–second “precursor pulse” is followed by a brief near-disappearance in the X-ray band as the photosphere expands to $\gtrsim10^{3}$ km before receding into the main burst—so-called superexpansion (SE) bursts \citep{1978ApJ...221L..57H,1984ApJ...276L..41T,1984ApJ...277L..57L}. Such events have been identified in several LMXBs, including 4U~1820$-$30, M15~X$-$2, 2S~0918$-$549, 4U~0614$+$091, and 4U~1722$-$30 \citep{2002ApJ...566.1045S,1990PASJ...42..633V,2011A&A...525A.111I,2010A&A...514A..65K,2003A&A...399..663K,2010A&A...520A..81I}. Most sources showing superexpansion are hydrogen-poor ultracompact X-ray binaries (UCXBs) or candidates, consistent with helium-dominated fuel and radiation-driven winds/outflows during super-Eddington phases \citep{2010A&A...520A..81I,2012A&A...547A..47I}. Recently, \emph{NICER} again observed 4U~1820$-$30 and reported moderate, strong, and even extreme PRE episodes—including one superexpansion case with $r_{\rm ph}>10^{3}$ km—during 2017--2023 \citep{2024A&A...683A..93Y}.

The source 2S~0918$-$549 is a prototypical UCXB candidate: it shows an unusually high X-ray/optical flux ratio and weak/absent H features in the optical, while high-resolution X-ray spectroscopy reveals an elevated Ne/O ratio indicative of a degenerate donor \citep{2004MNRAS.348L...7N,2003ApJ...599..498J}. Its likely orbital period is $17.4\pm0.1$ min from Gemini/GMOS time-series photometry, further supporting the UCXB nature \citep{2011ApJ...729....8Z}. Distance estimates from PRE “standard-candle’’ scalings lie in the $\sim4$–5.5 kpc range, depending on atmospheric composition and flux calibration \citep{2002A&A...392..885C,2005A&A...441..675I,2014A&A...568A..69I}. The long-term accretion rate is persistently at $\lesssim 1\%\,L_{\rm Edd}$, in line with the UCXB picture of low $\dot{M}$, helium accumulation, and long/intermediate-duration bursts \citep{2008ApJS..179..360G}. To date, \emph{BeppoSAX}/WFC has captured two long bursts from 2S~0918$-$549 (at least one with clear superexpansion), and \emph{RXTE}/PCA recorded five bursts between 2000–2008, including an event with an exceptionally short $\sim$30–40 ms precursor consistent with extreme expansion and fast outflows \citep{2002A&A...392..885C,2011A&A...525A.111I,2014A&A...568A..69I,2014A&A...565A..25I}. These hallmark features (precursor + “valley’’ + recovery) match the defining phenomenology of superexpansion bursts \citep{2010A&A...520A..81I}.

In this work, we apply the direct cooling tail method to constrain, for the first time, the mass and radius of 2S~0918$-$549. The paper is organized as follows. In Section~2, we summarize the RXTE observations of 2S~0918$-$549 and analyze the burst spectroscopy. In section~3, we introduce the direct-cooling method. In Section~4, we apply the direct-cooling method to estimate the neutron-star mass and radius. Section~5 presents a discussion of the results, and Section~6 gives our conclusions.

\section{Observations and Data Reduction} 
The Rossi X-ray Timing Explorer (RXTE), launched in 1995, carried three instruments---the Proportional Counter Array (PCA), the High Energy X-ray Timing Experiment (HEXTE), and the All-Sky Monitor (ASM)---to study rapid X-ray variability from neutron stars, black holes, pulsars, and active galactic nuclei \citep{1993A&AS...97..355B,1998ApJ...496..538R,1996ApJ...469L..33L}. In 2008, the PCA detected a Type I X-ray burst from 2S 0918$-$549 (RXTE ObsID 93416$-$01$-$05$-$00) \citep{2011A&A...525A.111I}. The PCA comprises five co-aligned, non-imaging proportional-counter units (PCUs) with an effective area of $\sim6.5\times10^{3}\,\mathrm{cm^{2}}$ at 6 keV, a bandpass of 2-60 keV, a typical spectral resolution of $\sim$18-20\% at 6~keV, and a collimated field of view of $2^{\circ}\times2^{\circ}$ (full width at zero response) \citep{2006ApJS..163..401J}.
This burst was an unusual long event: the light curve started with a $0.04$~s precursor, the main burst rose $\simeq1.2$~s later, and the emission maintained near-Eddington fluxes for the initial $\approx77$~s, all indicative of photospheric superexpansion \citep{2011A&A...525A.111I}. Superexpansion PRE bursts are rare and exhibit an extreme expansion phase during which the photosphere is pushed to $r_{\rm ph}\gtrsim10^{3}$~km for a few seconds, followed by a sustained near-Eddington ``moderate'' expansion phase with $r_{\rm ph}\sim30$~km \citep{2010A&A...520A..81I}.

\subsection{Persistent Emission}
\begin{figure}[t]
  \centering
  \includegraphics[
    height=0.7\columnwidth,
    keepaspectratio,
    angle=-90,
    origin=c
  ]{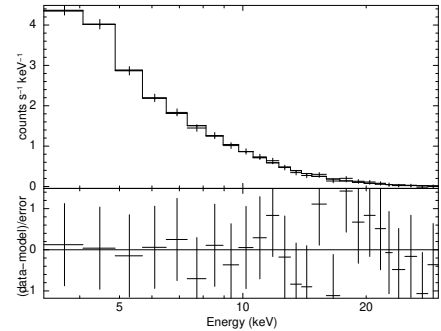}
  \caption{The persistent emission spectrum of 2S 0918--549 (observation ID: 93416-01-05-00).
  The reduced $\chi^2$ is 0.45.}
  \label{fig:spec-2s0918}
\end{figure}

For the PRE burst in 2S~0918$-$549, we extracted the persistent (preburst) spectrum from the 100~s immediately prior to burst onset using PCA Standard-2 data, which include both source emission and instrumental background \citep{2006ApJS..163..401J}. The instrumental background was estimated with \texttt{pcabackest}; because the persistent count rate was $<40~\mathrm{counts~s^{-1}~PCU^{-1}}$, we adopted the faint (L7) background model. The persistent emission in the 3-30 keV band is well described by an absorbed power law (\texttt{tbabs*powerlaw}) in XSPEC \citep{1996ASPC..101...17A,2000ApJ...542..914W}. The details of the best-fit parameters can be found in Table ~\ref{tab:preburst_spec}. Figure~\ref{fig:spec-2s0918} shows the preburst persistent spectrum and fit residuals; the fit yields a reduced $\chi^{2}_{\nu}=0.45$ for 23 degrees of freedom.

\begin{table}
\caption{The results of the spectral fit of the pre-burst spectrum with tbabs*powerlaw.
Errors are for 90\% confidence.
}
\label{tab:preburst_spec}
\centering
\begin{tabular}{ll}
\hline\hline
Model & power law \\
\hline
$N_{\rm H}$                    & $(3.0 ) \times 10^{21}\ {\rm cm^{-2}}$ \\
$\Gamma$                       & $2.23 \pm 0.03$ \\
$\chi^2_\nu$                   & $0.54\ (23\ {\rm dof})$ \\
Unabs.\ 0.1--200 keV flux      & $(5.6 \pm 0.05)\times10^{-10}\ {\rm erg\,cm^{-2}\,s^{-1}}$ \\
\hline
\end{tabular}
\end{table}

\subsection{Fitting the Burst Spectra}
We filtered the RXTE/PCA event data using the standard screening criteria. Although two proportional counter units (PCUs 0 and 2) were on during the burst, we excluded PCU~0 because its propane veto layer was lost in 2000, and we therefore analyzed only PCU~2, combining all xenon detector layers  \citep{2006ApJS..163..401J}. Good time intervals (GTIs) required elevation angle $\mathrm{ELV}>10^\circ$ and pointing offset $\mathrm{OFFSET}<0.02^\circ$; following the RXTE team's recommendations for faint sources, we also applied $(\mathrm{TIME\_SINCE\_SAA}>30~\mathrm{min}\ \mathrm{or}\ <0)$ and $\mathrm{ELECTRON2}<0.1$ cuts. 

Background spectra (particle $+$ cosmic) were generated with \texttt{pcabackest} over the same GTIs, using the faint (L7) model and a minimum time step of 16s. Because this burst exhibits a very short precursor and then remains near the Eddington flux (superexpansion/shell ejection), the earliest photons are expected to be largely out of the PCA bandpass; we therefore ignored the precursor spectrum and began the time-resolved burst spectroscopy 3s after the burst onset, extracting 1s spectra to maintain roughly uniform signal-to-noise across the sequence \citep{2011A&A...525A.111I,2010A&A...520A..81I}.
We generated PCA response matrices for each spectrum with \texttt{pcarsp} and applied the standard PCA deadtime correction following the RXTE Cook Book. Spectra were grouped to a minimum of 15 counts per bin (to ensure the validity of $\chi^2$ statistics) using \texttt{grppha}. To account for calibration uncertainties, we added a 0.5\% systematic error per channel as recommended by the PCA team.

To avoid the effects of burst-driven changes to the persistent emission during the expansion/precursor phase \citep{2004ApJ...602L.105B,2011A&A...525A.111I}, we exclude the precursor/strong-expansion interval. Following the procedure of \citet{2011A&A...525A.111I}, we fit each time-resolved spectrum during the burst with \texttt{tbabs*(bbodyrad+pow)}, where the power-law component represents the pre-burst persistent emission and its parameters are fixed at the pre-burst best-fit values during the burst fits.

\begin{figure*}[t]
  \centering
  \includegraphics[width=0.8\textwidth]{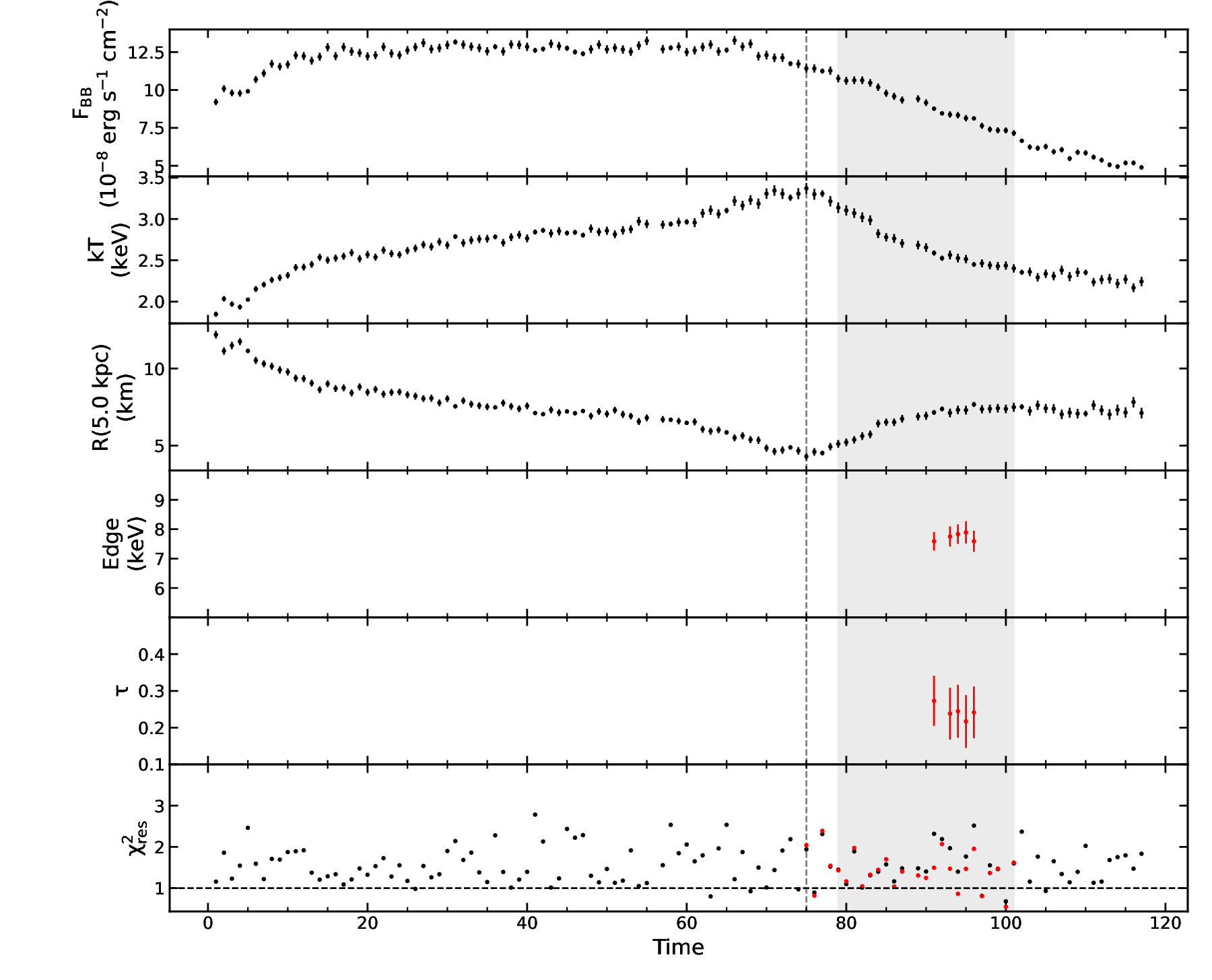}
      \caption{From the top to the bottom, the panels show the bolometric flux, the blackbody temperature and radius (assuming a distance of 5 kpc), the absorption-edge energy, the absorption optical depth and the reduced $\chi^{2}$. The black dots are the fitting results using the \texttt{TBABS*(BBODYRAD+POW)} model, and the red dots are from the \texttt{TBABS*(BBODYRAD*EDGE+POW)} model. The shaded region highlights the range when the flux is between \(0.95 F_{\rm td} - 0.6 F_{\rm td}\). The vertical dashed line marks the touchdown position.}
  \label{fig:spec-figure2}
\end{figure*}

The bbodyrad component yields the evolution of $kT_{\rm bb}$ and the blackbody normalization $K_{\rm bb}$; we fixed the absorption to $N_{\rm H}=3\times10^{21}\,\mathrm{cm^{-2}}$, consistent with previous studies of 2S~0918$-$549 \citep{2003ApJ...599..498J,2005A&A...441..675I,2000ApJ...542..914W}. Unabsorbed (0.1--200~keV) fluxes were computed with \texttt{cflux} wrapped around the burst component, and parameter uncertainties are quoted at 90\% confidence. The best-fit spectral parameters for different epochs are listed in Table~\ref{tab:burst_spectra}, including $kT_{\rm bb}$, $R_{\rm bb}$, the unabsorbed 0.1--200 keV flux of the burst component, and the fit quality ($\chi^{2}/{\rm dof}$). Using the same fixed $N_{\rm H}$ and the same unabsorbed 0.1--200 keV flux definition, we compared our burst spectral evolution with \citet{2011A&A...525A.111I} and find that the overall values (e.g. $kT_{\rm bb}$, $R_{\rm bb}$ and $F_{\rm bol}$) are consistent. The blackbody model provides good fits throughout the burst (see Fig.~\ref{fig:spec-figure2}).
In addition, we considered the reflection models such as \texttt{relxillNS\footnote{\url{http://www.sternwarte.uni-erlangen.de/~dauser/research/relxill/}}} to account for accretion-disk reflection; however, the resulting best-fit parameters were unphysical.

\begin{table}
\begin{flushleft} 

\caption{Spectral-fit results for three representative time intervals of the burst from 2S 0918–549 with tbabs*(bbodyrad+pow)\tablenotemark{1}.}
\label{tab:burst_spectra}

\small
\setlength{\tabcolsep}{5pt}        
\renewcommand{\arraystretch}{1.15} 
\protect\revone{
\begin{tabular}{@{}lccc@{}} 
\hline
Parameters & Contraction & Touchdown & Cooling tail \\
\hline
\shortstack[l]{Time interval\\(+54504 MJD)} & 61--62 & 73--74 & 85--86 \\
\hline
$N_{\rm H}$ ($10^{22}\ \rm cm^{-2}$)  & $0.3$ & $0.3$ & $0.3$ \\
$kT_{\rm bb}$ (keV)                  & $3.10^{+0.05}_{-0.05}$ & $3.37^{+0.06}_{-0.06}$ & $2.70^{+0.05}_{-0.05}$ \\
$R_{\rm bb}$\tablenotemark{2} (km)   & $5.69^{+0.17}_{-0.17}$ & $4.53^{+0.14}_{-0.14}$ & $6.36^{+0.21}_{-0.21}$ \\
$\chi^2/{\rm dof}$                   & 47/23 & 47/25 & 30/22 \\
\shortstack[l]{$F_{\rm bol}$\tablenotemark{3}\\($10^{-8}\,\mathrm{erg\,cm^{-2}\,s^{-1}}$)}
                                    & $13.0\pm0.2$ & $11.4\pm0.2$ & $9.34\pm0.2$ \\
\hline
\end{tabular}
}

\tablenotetext{1}{The pow component was fixed to the pre-burst persistent-emission best-fit parameters during the burst spectral fits.}
\tablenotetext{2}{Assuming a source distance of $D = 5$ kpc.}
\tablenotetext{3}{Unabsorbed flux in the 0.1--200 keV energy range. Errors are 90\% confidence.}

\end{flushleft}
\end{table}

From the time-resolved spectral fits, we measure a touchdown flux of \(F_{\rm td}=11.42\times10^{-8}\,\mathrm{erg\,s^{-1}\,cm^{-2}}\) (all fluxes are bolometric, 0.1--200~keV, in cgs units, as described above). Owing to the strong PRE and the near-constant bolometric flux during the PRE phase, the burst is regarded as having reached the Eddington limit \citep{2011A&A...525A.111I}. The ratio of the peak flux to the touchdown flux is \(f \equiv F_{\rm pk}/F_{\rm td}\approx1.16\); since \(f \lesssim 1.6\) is expected for non–high-inclination systems (with larger values predominantly found for high-inclination dippers), this points to a relatively low inclination, consistent with prior estimates for 2S~0918$-$549 \citep{2008MNRAS.387..268G,2011ApJ...729....8Z}. This is also consistent with the results reported by \citet{2011A&A...525A.111I}, which imply $f \approx 1.10$--$1.12$.
 Furthermore, using our fit to the pre-burst spectrum (Table~\ref{tab:preburst_spec}) 
we measure a persistent unabsorbed 0.1--200 keV flux of 
$(5.6 \pm 0.05)\times10^{-10}\ {\rm erg\,cm^{-2}\,s^{-1}}$, 
which corresponds to $\approx 0.5\%$ of the bolometric burst peak flux obtained in our analysis. 
This value is fully consistent with the $5.5\times10^{-10}\ {\rm erg\,cm^{-2}\,s^{-1}}$ reported by 
\citet{2011A&A...525A.111I}. This implies a very low mass accretion rate, under which the accretion flow is not expected to significantly affect the neutron-star atmosphere or the spectral evolution during the burst; in particular, the event satisfies the criterion of \citet{2016EPJA...52...20S} that the accretion luminosity be $\lesssim 5\%$ of the Eddington luminosity.

We also searched for variations in the persistent emission during the burst by introducing the variable-persistent (\(f_a\)) component and applying an F-test to assess whether the extra parameter is required. The \(f_a\) model did not yield a significantly improved reduced \(\chi^2\), implying that any accretion-rate enhancement during the burst makes a negligible contribution to the burst spectra \citep{2013ApJ...772...94W,2015ApJ...801...60W}.

Heavy-element absorption edges (e.g., near the Fe-peak) have been detected during the expansion phases of some superexpansion PRE bursts and are interpreted as nuclear-burning ashes carried into the line of sight by burst-driven winds/shell ejection \citep{2010A&A...520A..81I,2006ApJ...639.1018W}. To assess the impact of heavy elements on the color-correction factor \(f_{\rm c}\), we compared a ``simple'' model without edges (\texttt{tbabs*bbodyrad}) to a ``full'' model that adds, in each time bin, a photoionization edge with free energy \(E\) and optical depth \(\tau\). We performed this comparison over the touchdown–\(0.6\,F_{\rm td}\) cooling interval using the Bayesian Information Criterion (BIC; \citealt{1978AnSta...6..461S}) and the Akaike Information Criterion (AIC; \citealt{1974ITAC...19..716A}). In the Gaussian/high-count limit where \(-2\ln\mathcal{L}\simeq\chi^{2}\), these are \(\mathrm{BIC}=\chi^{2}+k\ln n\) and \(\mathrm{AIC}=\chi^{2}+2k\), with \(k\) the number of free parameters and \(n\) the number of data points. For a joint fit to 26 time bins, the simple model yields \(\chi^{2}/\mathrm{dof}=912.94/590\), while the full model gives \(767.29/538\). Thus \(\Delta\chi^{2}=-145.65\) and the added parameter count is \(\Delta k=52\). Taking \(n\) to be the total number of spectral channels (so that \(\ln n\simeq 6.47\)), we find \(\Delta \mathrm{BIC}\approx +1.90\times 10^{2}\)\, which strongly favors the simpler model (smaller BIC is preferred). The AIC difference, \(\Delta \mathrm{AIC}\approx -4.17\times 10^{1}\)\, formally favors the full model, but only by spending a large number of extra degrees of freedom; moreover, in most time bins the edge parameters are poorly constrained. Considering both the statistical criteria and these physical diagnostics, we conclude that the simple blackbody model is sufficient for this burst’s cooling tail, and no metal-enriched atmosphere is required to explain the data \citep[see also the discussion of superexpansion-edge phenomenology in][]{2010A&A...520A..81I}.

\section{METHOD} 

In our analysis we follow the approach of \citet{2012A&A...545A.120S} to compute theoretical neutron‐star atmosphere models and to relate them to observed burst spectra. We consider non‐rotating neutron stars in hydrostatic equilibrium under the plane‐parallel approximation. 

We repeat here Eqs.~(1)–(27) of \citet{2017MNRAS.466..906S} for completeness (more details can be found in~\citet {2017MNRAS.466..906S}). For a non-rotating NS, the surface gravity is \citep{2012A&A...545A.120S,2015A&A...581A..83N}
\begin{equation}
 g = \frac{GM}{R^2}\,(1+z),
\end{equation}
and the surface effective temperature is given by
\begin{equation}
 \sigma_{\rm SB}T_{\rm eff}^4 = \frac{L}{4\pi R^2}.
\end{equation}
where $1+z=(1-R_{\rm S}/R)^{-1/2}$ is the redshift factor and $R_{\rm S}=2GM/c^2$ is the Schwarzschild radius of the neutron star and $\sigma_{\rm SB}$ is the Stefan–Boltzmann constant. The critical (local) Eddington luminosity for spherically symmetric emission from a non‐rotating NS surface is given by \citet{1993SSRv...62..223L} as 
\begin{equation}
 L_{\rm Edd} = \frac{4\pi\,G\,M\,c\,(1+z)}{\kappa_{\rm T}} = {4\pi}R^2\sigma_{\rm SB}T_{\rm edd}^4.
\end{equation}
where $\kappa_{\rm T}=0.2\,(1+X)\,\mathrm{cm^2\,g^{-1}}$ is the Thomson electron‐scattering opacity and $X$ is the hydrogen mass fraction of the atmospheric plasma.  Due to general‐relativistic effects, the luminosity $L_{\infty}$, the effective temperature $T_{\rm eff,\infty}$, and the apparent radius $R_{\infty}$ observed at infinity differ from the surface values $L$, $T_{\rm eff}$, and $R$ according to \citet{1993SSRv...62..223L}:
\[
R_{\infty} = R\,(1+z),\quad
T_{\rm eff,\infty} = \frac{T_{\rm eff}}{1+z},\quad
L_{\infty} = \frac{L}{(1+z)^2}.
\]
The observed Eddington luminosity, Eddington flux, and the Eddington temperature at infinity are then 
\begin{equation}
 L_{\rm Edd,\infty} = \frac{4\pi GMc}{\kappa_{\rm T}}\,\frac{1}{(1+z)},
\end{equation}
\begin{equation}
F_{\rm Edd,\infty} = \frac{L_{\rm Edd,\infty}}{4\pi D^2} = \frac {GMc}{\kappa_{\rm T} D^2}\,\frac{1}{(1+z)}.
\end{equation}
\begin{equation}
T_{\rm Edd,\infty}
= \left( \frac{g\,c}{\sigma_{\rm SB}\,\kappa_{\rm T}} \right)^{1/4}\,\frac{1}{1+z}.
\end{equation}
These relations follow the standard treatment adopted in \citet{2017MNRAS.466..906S}. Observationally, neutron star burst spectra are well fit by a thermal blackbody spectrum characterized by a color temperature $T_{\rm BB}$ and a normalization $K=R_{\rm BB}^2/D^2$ \citep{2008ApJS..179..360G,2012ApJ...747...76G,2015ApJ...801...60W}:
\begin{equation}
F_{\rm BB} = \sigma_{\rm SB}\,T_{\rm BB}^4\,K.
\end{equation}
the observed bolometric flux is determined as 
\begin{equation}
F_{\rm \infty} = \sigma_{\rm SB}\,T_{\rm \infty}^4\,\Omega.
\end{equation}
where
\begin{equation}
\Omega = \frac{R_\infty^2}{D^2}
\end{equation}
During an X-ray burst, the intense radiation traverses the hot plasma atmosphere, where electron scattering (Compton effect) and photon absorption dilute the intrinsic blackbody spectrum \citep{1984PASJ...36..551E}. To derive the bolometric burst flux, a color-correction factor $f_{\rm c}=T_{\rm BB}/T_{\rm eff}$ should be applied
\begin{equation}
F_E \approx w\,B_E\bigl(f_{\rm c}T_{\rm eff}\bigr).
\end{equation}
where $B_E(T)$ is the Planck function at temperature $T$,
$f_c \equiv T_{\rm BB}/T_{\rm eff}$ is the color-correction factor,
and $w$ is the dilution factor. Integration over all energies yields
\begin{equation}
\int_{0}^{\infty} w\,\pi\,B_E\!\left(f_c\,T_{\rm eff}\right)\,{\rm d}E= w\,f_c^{4}\,\sigma_{\rm SB}\,T_{\rm eff}^{4}.
\end{equation}
Thus, we can relate the observed blackbody flux to the relative luminosity as 
\begin{equation}
 F_{\rm \infty} = (w f_{\rm c}^4)^{\rm -1}\,F_{\rm BB}.
\end{equation}
Hence,
\begin{equation}
 F_{\rm BB} = (w f_{\rm c}^4)\,l\,F_{\rm Edd}.
\end{equation}
Therefore, the relation between the observed blackbody normalization $K$, the
neutron-star radius and distance (through $\Omega = R_\infty^2/D^2$), and the
spectral fitting parameters $(w,f_c)$ becomes
\begin{equation}
  (w f_c^{4})^{-1}\,T_{\rm BB}^{4}\,K = T_\infty^{4}\,\Omega,
\end{equation}
or
\begin{equation}
  K= w\,\Omega.
\end{equation}

The theoretical dependence of \(f_{\rm c}\) on the relative luminosity \(l\), surface gravity \(\log g\), and composition (\citealt{2011A&A...527A.139S}; \citealt{2012A&A...545A.120S}; \citealt{2015A&A...581A..83N}) provides model curves of \(K\) versus \(F_{\rm BB}\). Following the implementation of the cooling-tail method in \citet{2017MNRAS.466..906S}, by interpolating these curves over a grid of \(\log g\) values and fitting to the observed \(K\)--\(F_{\rm BB}\) data, one obtains the best-fit \(F_{\rm Edd}\) and \(\Omega\).  These measurements translate into confidence regions in the \(M\)--\(R\) plane.  This cooling-tail method has been applied successfully to low-mass X-ray binaries such as SAX~J1810.8–2609 and GRS~1747–312 \citep{2017MNRAS.466..906S,2018ApJ...866...53L,2017A&A...608A..31N}.

\section{Mass--Radius Constraints}
We constrain the neutron-star mass and radius in 2S~0918$-$549 using the direct cooling-tail method \citep{2017MNRAS.466..906S,2017A&A...608A..31N}. We evaluate a grid in the $(M,R)$ plane with
$M\in[1.0,3.0]\,M_{\odot}$ (step $0.01\,M_{\odot}$) to be consistent with the NS formation scenarios and the observational pulsar mass measurements/constraints~\citep{2002RvMP...74.1015W,2014MNRAS.442.3777P}, and $R\in[3,18]$ km (step $0.01$ km). Grid points that violate the causality condition $R>2.9\,GM/c^{2}$ \citep{2007PhR...442..109L} are discarded. For each accepted $(M,R)$ we compute the surface gravity and gravitational redshift,
and the apparent radius $R_{\infty}=R(1+z)$. We adopt atmosphere grids for two compositions: a metal-rich model \citep{2017A&A...608A..31N} and a pure-helium model \citep{2017MNRAS.466..906S}. In these grids we use $\log g$  spanning $13.7$--$14.9$ in steps of $0.1$ and obtain intermediate values by linear interpolation.

In a given atmospheric composition and surface gravity $g$, the atmosphere model provides the dilution factor $w(\ell)$ and the color-correction factor $f_c(\ell)$ as functions of the relative luminosity $\ell \equiv F/F_{\mathrm{Edd},\infty}$. During the cooling tail, the observables derived from blackbody fits are the blackbody normalization $K_{\mathrm{bb}}$ and the bolometric blackbody flux $F_{\mathrm{BB}}$, for which the model predicts $K_{\mathrm{bb}} = w(\ell)\,(R_{\infty}/D)^{2}$ and $F_{\mathrm{BB}} = w(\ell)\,f_c(\ell)^{4}\,\ell\,F_{\mathrm{Edd},\infty}$. \revone{Following \citet{2017A&A...608A..31N}, we restrict the cooling-tail fit to bins with flux ratio \(F/F_{\rm td}\in[0.6,0.95]\); for 2S~0918$-$549 this corresponds to  \(t\simeq78\)--99~s after touchdown.} We fit the data from $0.95\,F_{\mathrm{td}}$ to the $0.6\,F_{\mathrm{td}}$ level by minimizing the sum of squared, uncertainty-weighted distances from the model curve $(K(\ell),\,F(\ell))$, \revone{as in Eq. (28) of \citet{2017MNRAS.466..906S}, (see also \citep{Deming2011}):}:
\begin{equation}
\begin{aligned}
\chi^{2} &= \sum_{i=1}^{N_{\rm obs}}
\left[
\frac{\big(w\,\Omega-K_{i}\big)^{2}}{\sigma_{K,i}^{2}}+
\frac{\big(w\,f_{c}^{4}\,\ell\,F_{\rm Edd}-F_{i}\big)^{2}}{\sigma_{F,i}^{2}}
\right].
\end{aligned}
\label{eq:chi}
\end{equation}
with $(K_{i},F_{i})$ and $(\sigma_{K,i},\sigma_{F,i})$ the observed values and the corresponding uncertainties in time bin $i$. For each candidate pair $(M, R)$, we linearly interpolate the atmosphere model grid in surface gravity $\log g$ to obtain the $w f_{\rm c}^4$--$\ell$ relation;
we compute $\chi^2$ via Eq.~(\ref{eq:chi}) and obtain the minimum $\chi^2$ with the best fit $D$.

\begin{figure}[t]
  \centering
  \includegraphics[width=\columnwidth]{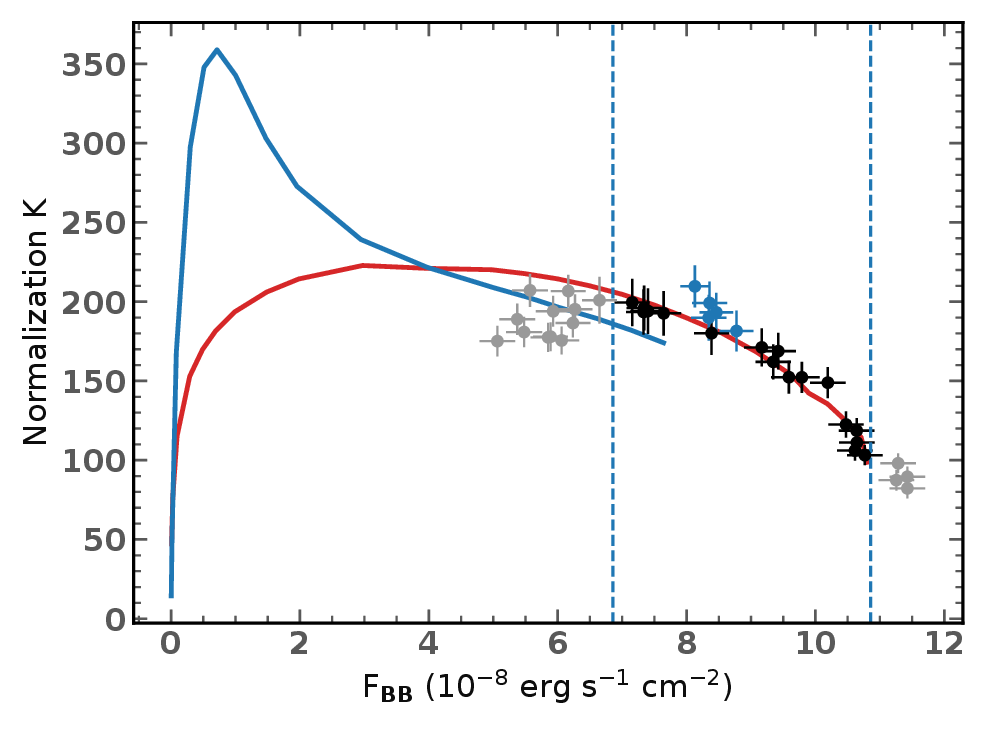}
  \caption{Dependence of blackbody normalization on flux for the PRE burst. Data are shown as circles with error bars. The solid red curve denotes the best-fit model for a pure He atmosphere, while the solid blue curve shows the best-fit metal-rich atmosphere model. Only the black points between the vertical dashed lines are used in the fit. Blue points indicate spectra with significant absorption edges and are excluded from the fit. Grey points denote data with fluxes outside the \(0.95 F_{\rm td} - 0.6 F_{\rm td}\) range, which are not used in the fit.}
  \label{fig:spec-figure3}
\end{figure}

\begin{figure}[t]
  \centering
  \includegraphics[width=\columnwidth]{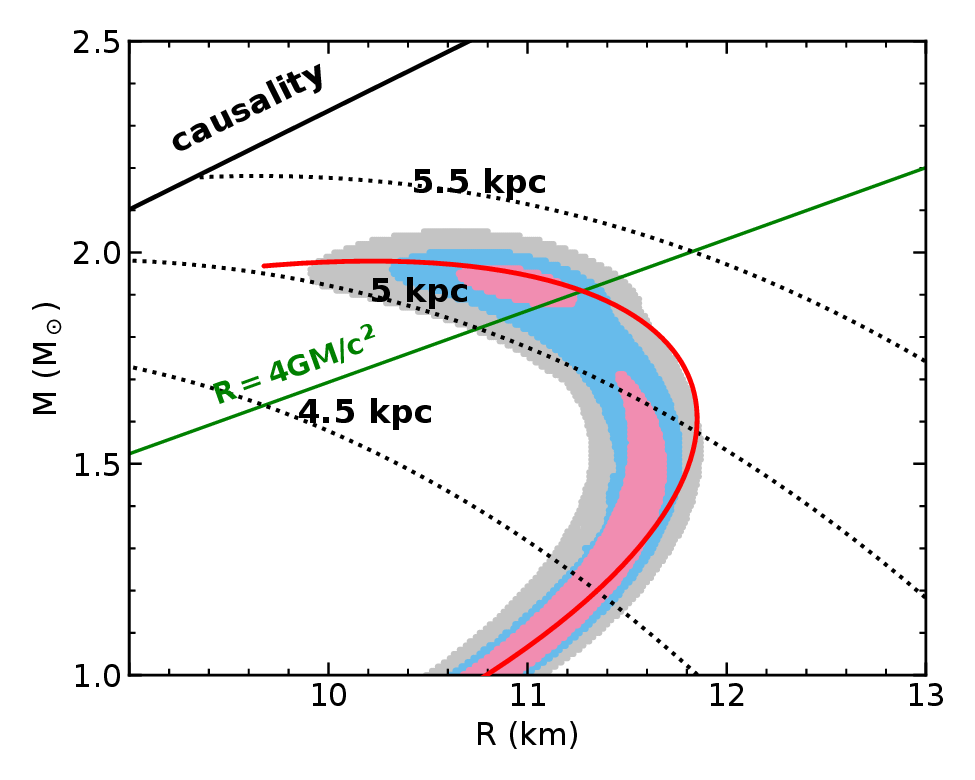}
  \caption{Mass--radius constraints for the neutron star in 2S~0918$-$549 using a pure-helium atmosphere model. The pink, blue, and gray contours show the 68\%, 90\%, and 99\% confidence regions, respectively. The best fit yields $\chi^{2} = 18.12$ for 14 dof. The red solid line correspond to the constant Eddington temperature, $T_{\mathrm{Edd}}$, given by the equation of the best-fit pure-helium model. The solid green line correspond to the critical radius $R = 4GMc^2$. The black dotted curves correspond to constant distances of 4.5, 5, and 5.5\,$\mathrm{kpc}$ for the pure-helium model. }
  \label{fig:spec-figure4}
\end{figure}

The global minimum of $\chi^{2}$ yields the best-fit parameters. Following the statistical treatment adopted by \citet{2017MNRAS.466..906S}, 
 we construct confidence regions in the $(M,R)$ plane using 
$\Delta\chi^2(M,R) = \chi^2(M,R) - \chi^2_{\min}$. 
For two parameters of interest, $M$ and $R$, the contours at 
$\Delta\chi^2 = 2.30,\,4.61,$ and $9.21$ correspond to the 68\%, 90\%, 
and 99\% joint confidence levels, respectively. As discussed by \citet{2017MNRAS.466..906S}, this simple $\chi^2$ approach tends to underestimate the parameter uncertainties compared with the more robust likelihood method of \citet{2014MNRAS.442.3777P}. However, the centroid of the constraints in the $(M,R)$ plane is not expected to be significantly affected. We do not analyse the part of the lightcurve affected by the rapid fluctuations in flux (after 120 s) discussed by \citet{2011A&A...525A.111I}, which the authors suggest are likely caused by perturbations to the disc due to the burst flux. We remove bins that show clear absorption edges to minimize the impact of metal ashes on the atmosphere modeling \citep{2010A&A...520A..81I,2006ApJ...639.1018W}. Among the tested compositions, the pure He atmosphere provides the best fit in Fig.~\ref{fig:spec-figure3}, with \(\chi^{2}/\nu=18.12/14\), whereas metal-rich models do not reproduce the data. In Fig.\ref{fig:spec-figure4}, we map \(\Delta\chi^{2}\) to confidence regions in the \((M,R)\) plane using the standard two-parameter thresholds.

From the above direct cooling tail calculations, we find that the best-fitting model  corresponds to
\(\log g=14.01\) for pure He atmosphere. The associated best-fit parameters are
\(F_{\rm Edd,\infty}=10.25 \times 10^{-8}\ {\rm erg\ cm^{-2}\ s^{-1}}\)
and
\(\Omega=923.91\,({\rm km}/10\,{\rm kpc})^{2}\). Using the two fitting parameters in this procedure, the apparent Eddington flux
\(F_{\rm Edd,\infty}\) and the angular factor \(\Omega\), we yield the Eddington temperature (see also Eq.(27) of~\citet{2017MNRAS.466..906S}):
\[
T_{\rm Edd,\infty}
= 9.81\ \mathrm{keV}\ \left(\frac{F_{\rm Edd,-7}}{\Omega}\right)^{1/4},
\]
where \(F_{\rm Edd,-7}\equiv F_{\rm Edd}/10^{-7}\ \mathrm{erg\ s^{-1}\ cm^{-2}}\)  \citep{2017A&A...608A..31N,2011A&A...527A.139S}.
The constant Eddington temperature $T_{\rm Edd,\infty}=1.79\,{\rm keV}$ corresponding to the best fitting parameters is plotted in Fig.~\ref{fig:spec-figure4} (red solid curve), which is consistent with the direct cooling-tail confidence region.
Using $\Omega = R_\infty^2/D^2$ with $R_\infty = R(1+z)$, we also show iso-distance curves (dotted lines), which provide a visual indication of the distance range implied by our best-fitting $\Omega$.

We note that the $M$--$R$ posterior distribution extends to, and partially
crosses, the critical-radius relation $R = 4GM/c^{2}$; this behavior is
consistent with the well-known posterior structure of the cooling-tail
method, where the joint mass--radius constraints often exhibit a bimodal
or banana-shaped morphology~\citep{2015ApJ...810..135O,2011ApJ...742..122S,2014MNRAS.442.3777P,2016A&A...591A..25N,2017MNRAS.466..906S}. 

\section{Discussion}
We performed time-resolved spectroscopy of an intermediate-duration ($\sim40$~min) thermonuclear X-ray burst from 2S~0918$-$549 that exhibits extreme photospheric radius expansion. The two stage evolution of this burst (superexpansion
and moderate expansion) is consistent with SE bursts reported in 4U~0614+091, 4U~1722$-$30, and 4U~1820$-$30 \citep{2010A&A...514A..65K,2003A&A...399..663K,2002ApJ...566.1045S,2010A&A...520A..81I}. The cooling tail is, in bulk, well described by a pure-helium atmosphere. We also find tentative evidence for absorption edges after touchdown, although the signal is modest and we do not attempt a detailed identification of the responsible species.

\citet{2011A&A...525A.111I} analysed achromatic fluctuations of the light curve after t = 120 s. Since these are likely due to interference by a perturbed accretion disc \citep{2011A&A...525A.111I} they are excluded by our analysis. 

\subsection{M–R of the NS in 2S 0918-549}

Through cooling-tail plus atmosphere-model fitting, we obtain \(M = 1-2\,M_\odot\) and \(R = 9.7-11.9\,\mathrm{km}\) at 99\% confidence level. The $M$--$R$ contours in Fig.~5 show that, at the $1\sigma$ confidence level, four gravity-bound EOSs and four self-bound EOSs are all acceptable, indicating that, with the present data from this single bursting source, several distinct EOSs remain compatible with the confidence region. 

It is important to stress that our mass interval does not directly cover the low-mass regime of \(\lesssim 1\,M_\odot\); if such low-mass neutron stars exist and their radii are consistent with predictions of self-bound EOSs, their \(M\)–\(R\) relation would differ markedly from that of purely gravity-bound stars. Although several low-mass candidates (Her~X$-$1, SMC~X$-$1, 4U~1538$-$52, PSR~J1518+4904) have been proposed, robust radius measurements are still lacking, so discrimination among EOSs at the low-mass end remains limited by sample incompleteness and parameter degeneracies\citep{2005A&A...441..685V,2011ApJ...730...25R,2008A&A...490..753J}.
\begin{figure}[t]
  \centering
  \includegraphics[width=\columnwidth]{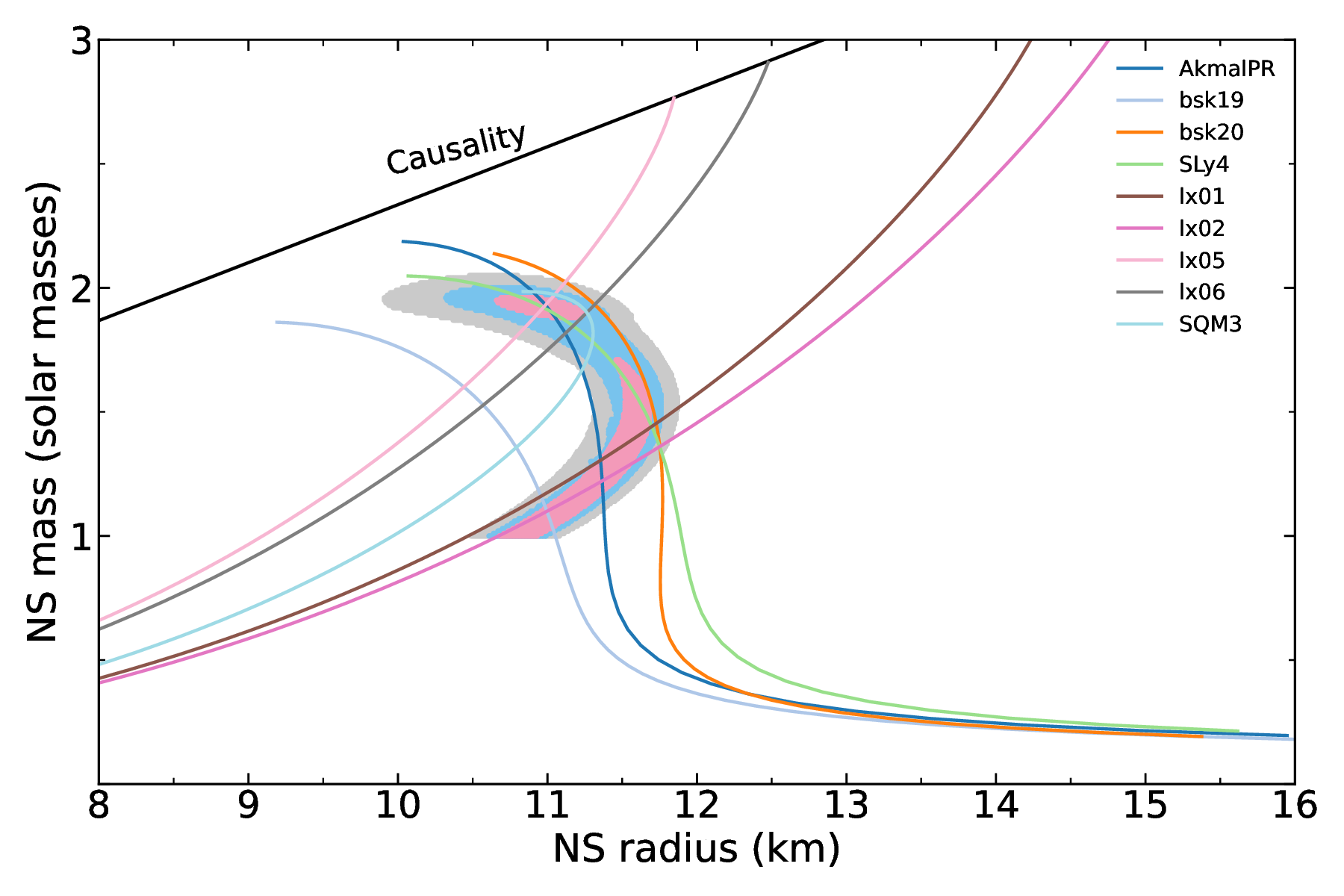}
  \caption{The pink, blue, and gray contours show the 68\%, 90\%, and 99\% confidence regions, respectively. Also shown are representative theoretical mass–radius relations: the unified Skyrme model \textsc{SLy4} \citep{2001A&A...380..151D}, the APR EOS 
  \citep{1998PhRvC..58.1804A}, the Brussels–Montreal unified EOS \textsc{BSk19}, \textsc{BSk20} (analytic representation from \citealt{2013A&A...560A..48P}), the self-bound strangeon star EOS \textsc{lx01}, \textsc{lx02}, \textsc{lx05} and \textsc{lx06} \citep{2009MNRAS.398L..31L} and the self-bound strange-quark–matter model \textsc{SQM3} \citep{2001ApJ...550..426L}.These curves are shown for comparison only; the shaded contours reflect the statistical uncertainties of our pure-He model fit.}
  \label{fig:spec-figure5}
\end{figure}
\subsection{Distance to 2S 0918-549}
Early optical identification with the ESO 2.2\,m telescope, under the assumption of typical intrinsic colors and absolute magnitudes for LMXBs, yielded a distance of \(\sim15\) kpc for 2S~0918$-$549, which was later considered too large \citep{1987A&A...172..167C}. Subsequent constraints from the X-ray burst method were tighter: the first RXTE-detected Type I burst gave an upper limit of \(\le 4.9\) kpc under the assumption that the peak does not exceed the Eddington limit \citep{2001ApJ...553..335J}; 
Furthermore, time-resolved spectral analysis 
of a PRE burst observed in 1999 
with \textit{BeppoSAX}/WFC 
yielded a distance estimate of $D \simeq 4.2 \,\mathrm{kpc}$, with a systematic uncertainty of $\sim 30\%$ \citep{2002A&A...392..885C,2003A&A...399..663K}. Further equating the peak burst flux to the Eddington luminosity and accounting for atmospheric composition yields representative distances of \(4.1\) kpc for H-rich and \(5.4\) kpc for He-rich/H-poor atmospheres \citep{2005A&A...441..675I,2014A&A...568A..69I}; many later works therefore adopted a “working distance” of \(\sim 5.4\pm0.8\) kpc, though values of \(\sim4.8\) kpc also appear \citep{2004MNRAS.354..355J}. In this study we apply the direct cooling-tail method, jointly fitting the \(K\)–\(F_{\rm BB}\) relation in the cooling phase while treating the distance \(D\) as a free parameter, and obtain \(D=4.1-5.3\) kpc. This value is consistent with the representative distance obtained from the ``peak $=$ Eddington'' calibration. This consistency reflects the compatibility of results despite differences in modeling assumptions: the traditional approach fixes canonical neutron-star parameters ($M=1.4\,M_\odot$, $R=12$ km) and infers the distance $D$ from $F_{\rm Edd}$, whereas the cooling-tail method jointly solves for $M$, $R$, and $D$ and explicitly accounts for atmosphere physics and color corrections, without assuming that the observed peak corresponds to the standard $L_{\rm Edd}$. Accordingly, our inferred distance agrees with previous measurements within the methodological and prior uncertainties. As an independent cross-check, we also consult Gaia DR3 \citep{GaiaCollab2023DR3},via the VizieR catalog I/352\footnote{\url{https://vizier.cds.unistra.fr/viz-bin/VizieR?-source=I/352}}: the counterpart’s parallax is $\varpi=0.2357 \pm 0.0606$ mas; using the EDR3 distance posterior of \citet{BailerJones2021AJ}, we obtain $r_{\rm geo}=4.31$ kpc (68\% credible interval: 3.31–5.79 kpc), consistent with the burst-based estimates and with the cooling-tail range derived here.

\section{Conclusion}
We analyzed an intermediate-duration thermonuclear burst from 2S~0918$-$549 observed with \emph{RXTE}, which exhibits extreme photospheric radius expansion. The cooling-tail spectra are well described by a pure-He atmosphere without requiring prominent heavy-element absorption edges, naturally pointing to a helium white-dwarf donor: long-term accumulation of He followed by ignition at critical conditions can account for the event. Applying the direct cooling-tail method, we obtained joint neutron-star mass–radius constraints (Fig.~\ref{fig:spec-figure5}): within the \(1\sigma\) region, representative families of both gravity-bound and self-bound equations of state (EOS) remain viable. These constraints complement the existence of \(\gtrsim 2\,M_\odot\) neutron stars—which already exclude overly soft EOS—while still allowing low-mass, small-radius solutions within current uncertainties.

The pure-He fit favors a distance range of $D=4.1-5.3\,\rm kpc$ (99\% confidence). Future improvements in atmosphere modeling, together with independent distance measurement, will substantially tighten the \(M\)–\(R\) confidence region and enhance EOS discrimination. Moreover, if the \(\sim 774.06\) Hz candidate burst oscillation reported by \emph{Swift}/BAT is confirmed\citep{2024FrASS..1177677L}, 
the spin can be incorporated into the atmosphere model and ISCO-related modeling, providing a more unified and better-constrained framework. In addition, combining the orbital period with the companion’s mass and radius \citep{2011ApJ...729....8Z} can provide insight into the close-binary evolutionary pathway.

\begin{acknowledgments}
We thank the anonymous referee for the constructive comments and suggestions. This work was supported by the Natural Science Foundation of Xinjiang Tianshan Talents program No.2024TSYCJU0001, the National Natural Science Foundation of China Nos. 12263006 and 12273030, the Natural Science Foundation of Xinjiang No. 2024D01C52, and the Major Science and Technology Program of Xinjiang Uygur Autonomous Region under grant No. 2022A03013-3.
\end{acknowledgments}

\bibliography{sample701}{}
\bibliographystyle{aasjournalv7}



\end{document}